\input amstex
\magnification 1100
\vsize 240 truemm
\hsize 160 truemm

\def\Nat{{I\!\! N}}
\def\Real{{I\!\! R}}
\def \E{{I\!\! E}}

\noindent
\centerline{\bf HAUSDORFF MOMENT PROBLEM VIA FRACTIONAL MOMENTS}
\bigskip\noindent
\bigskip\noindent
1. Introduction
\par\noindent
In Applied Sciences a variety of problems, formulated in terms of linear
boundary values or integral equations, leads to a Hausdorff moment
problem. Such a problem arises when a given sequence of real numbers may
be represented as the moments around the origin of non-negative measure,
defined on a finite interval, typically $[0,1]$. The underlying density
$f(x)$ is unknown, while its moments
$\mu_j=\int_0^1 x^j f(x)dx$, $j=0,1,2,...$,with $\mu_0=1$, are known.
Next, through a variety of techniques, for practical purposes $f(x)$ is
recovered by taking into account only a finite sequence
$\{\mu_j\}_{j=0}^M$. Such a process implies that $f(x)$ is well-characterized
by its first few moments.
On the other hand, it is well known that the moment problem becomes
ill-conditioned when the number of moments involved in the reconstruction increases [1,2].
In Hausdorff case, once fixed $(\mu_0,...,\mu_{M-1})$, the moment $\mu_M$
may assume values within the interval $[\mu_M^-,\mu_M^+]$, where [3]
$$
\mu_M^+-\mu_M^-\leq 2^{-2(M-1)}
\eqno(1.1)
$$
\noindent
If one considers the approximating density
$f_{M}(x)=\exp(-\sum_{j=0}^{M} \lambda_j x^j)$
by entropy maximization, constrained by the first $M$ moments
[4], then its entropy
$H[f_{M}]=-\int_0^1 f_{M}(x)\ln f_{M}(x)dx$ satisfies
$$
\lim\limits_{\mu_M\rightarrow \mu_M^\pm}H[f_{M}]=-\infty
\eqno(1.2)
$$
\noindent
Such a relationship is satisfied by any other distribution constrained by the same
first $M$ moments, since $f_{M}(x)$ has maximum entropy.
On the other hand $f(x)$ and $f_M(x)$ have the same first $M$ moments and as
a consequence, as we illustrate in section 3, the following relationship
holds
$$
I(f,f_{M})=:\int_0^1 f(x)\ln {f(x) \over f_{M}(x)} dx=
H[f_{M}]-H[f].
\eqno(1.3)
$$
\noindent
Here $H[f]$ is the entropy of $f(x)$, while $I(f,f_{M})$ is the Kullback-Leibler
distance between $f(x)$ and $f_M(x)$.
\par\noindent
Equations (1.1)-(1.3) underline once more the ill-conditioned nature of the moment problem.
\par\noindent
The ill-conditioning may be even enlightened
by considering the estimation of the parameters $\lambda_j$ of $f_M(x)$.
The $\lambda_j$ calculation leads to minimize a proper potential
function $\Gamma(\lambda_1,...,\lambda_M)$[Kesa 4], with
$$
\min_{\lambda_1,...,\lambda_M} \Gamma(\lambda_1,...,\lambda_M)=
\min_{\lambda_1,...,\lambda_M}\Bigl [\ln\Bigl (
\int_0^1 \exp(-\sum_{j=1}^{M} \lambda_j x^j)dx \Bigl )+
\sum_{j=1}^{M} \lambda_j \mu_j\Bigl ].
\eqno(1.4)
$$
\noindent
$f_M(x)$ satisfies the constraints
$$
\mu_j=\int_0^1 x^j\exp(-\sum_{k=0}^{M} \lambda_k x^k1)dx,\qquad j=0,...,M
\eqno(1.5)
$$
\noindent
Letting $\mu=(\mu_0,...,\mu_M)$ and $\lambda=(\lambda_0,...,\lambda_M)$,
(1.5) may be written as the map
$$
\mu=\phi(\lambda)
\eqno(1.6)
$$
\noindent
Then the corresponding Jacobian matrix, which is up to sign a Hankel matrix, has
conditioning number $\simeq (1+\sqrt{2})^{4M}/\sqrt{M}$ [5].
All the previous remarks lead to the conclusion that $f(x)$ may be efficiently
recovered from moments only if few moments are requested. In other terms,
$f(x)$ may be recovered from moments if its information content is spread
among first few moments.
\par\noindent
In this paper we are looking for a way to overcome the above-quoted
difficulties in recovering $f(x)$ from moments.
First of all, we assume  the infinite sequence of moments
$\{\mu_j\}_{j=0}^\infty$ to be known. Then, from such a sequence, we calculate
fractional moments
$$
E(X^{\alpha_j})=:\int_0^1 x^{\alpha_j} f(x)dx=
\sum_{n=0}^{\infty} b_n(\alpha_j) \mu_n,\;\; \alpha_j>0
\eqno(1.7)
$$
\noindent
where the explicit analytic espression of $b_n(\alpha_j)$ is given by (2.5).
Finally, from a finite number of fractional moments
$\{E(X^{\alpha_j})\}_{j=1}^M$, we recover
$f_{M}(x)=\exp(-\sum_{j=0}^{M} \lambda_j x^{\alpha_j})$
by entropy maximization [4]. The exponents $\{\alpha_j\}_{j=1}^M$
are chosen as follows
$$
\{\alpha_j\}_{j=1}^M:\;\; H[f_M]=\hbox{minimum}
\eqno(1.8)
$$
\noindent
The choice of $\{\alpha_j\}_{j=1}^M$, according to (1.8), leads to a
density $f_M(x)$ having minimum distance from $f(x)$, as stressed by (1.3).
\par\noindent
Remark. If the information content of $f(x)$ is shared among first
moments, so that ME approximant $f_M(x)$ represents an accurate approximation of $f(x)$,
then fractional moments may be accurately calculated by replacing $f(x)$
with $f_M(x)$. As a consequence, function $f_M(x)$ converges in entropy and then in $L_1-$norm to
$f(x)$ [6], and the error obtained replacing $f(x)$ with $f_M(x)$
$$
\mid E_f(X^{\alpha_j})-E_{f_M}(X^{\alpha_j})\mid\leq \int_0^1 x^{\alpha_j}
\mid f(x)-f_M(x)\mid dx\leq
$$
$$
\leq\int_0^1\mid f(x)-f_M(x)\mid dx
\leq\sqrt{2(H[f_M]-H[f])}
\eqno(1.9)
$$
\noindent
may be rendered arbitrarily small by increasing $M$ (inequalities in (1.9)
are proved in section 3).
\par\noindent
\bigskip\noindent
2. Fractional moments from moments
\par\noindent
Let $X$ a continuous random variable with density $f(x)$ on the support $[0,1]$,
with moments of order $s$, centered in $c, \,\,\, c \in \Real$
$$
\mu_s(c):=\E \left[(X-c)^s \right]=\int_0^1 (x-c)^s \, f(x) \, dx, \,\,\,\,\, s \in \Nat^*=\Nat\cup\{0\}.
\eqno(2.1)
$$
\noindent
and moments from the origin $\mu_s=:\mu_s(0)$ related to moments generically centered in $c$ through the relationship
$$
\mu_s=\sum_{h=0}^s {s \choose h} \,  c^{s-h} \, \mu_h(c), \,\,\,\,\, s \in \Nat^*.
\eqno (2.2)
$$
\noindent
It is well known the relationship similar to (2.2) which permits to calculate the (fractional) moment of order $s \in \Real^{+}$ 
(which replaces $\alpha_j$ for notational convenience as in (1.7) and (3.2)) involving all the central moments
of a given distribution about the point $c$.
\par\noindent
Firstly, by definition of noncentral moment of order $s$, we can write $\E(X^s)=\int_0^1 x^s f(x) dx$ and then,
by Taylor expansion of $x^s$ around $c$, where $c \in (0,1)$, we have
$$
\eqalign{
x^s &= \sum_{n=0}^\infty \left[ x^s \right]_{x=c}^{(n)} {{(x-c)^n} \over {n!}} \cr
                             &=\sum_{n=0}^\infty \left[  {s \choose n} \,\, n! \,\, x^{s-n} \right]_{x=c} {{(x-c)^n} \over {n!}} \cr
                             &=\sum_{n=0}^\infty {s \choose n} c^{s-n} (x-c)^n \cr}
\eqno(2.3)
$$
where $[k(x)]_{x=c}^{(n)}$ indicates the $n$-th derivative of the function $k(x)$ wrt $x$, evaluated at $c$.
\par\noindent
Taking the expectation on both sides of the last equation in (2.3), we get the required relationship
$$
\eqalign{
\E \left(X^s \right) &= \sum_{n=0}^\infty {s \choose n} c^{s-n} \E\left[(X-c)^n\right] \cr
                             &= \sum_{n=0}^\infty b_n \, \mu_n(c) \cr}
\eqno(2.4)
$$
where
$$
b_n={s \choose n} c^{s-n} , \,\,\, n \in \Nat^*
\eqno(2.5)
$$
represents the coefficient of the integral $n$-order moment of $X$ centered at $c$.
\par\noindent
The formulation of the $s$-order fractional moments as in (2.4) shows some numerical instabilities which depend 
on the structure of the relationship between $\mu_n(c)$ and $\E(X^s)$; these instabilities are related to the value
of the center $c$ and increase as the order of the central moments becomes high. In particular,
\smallskip
\par\noindent
\item{(a)} the numerical error $\Delta\E(X-c)^n$ due to the evaluation of $\E(X-c)^n$ in terms of noncentral
integral moments $\E(X^h)$, $h \leq n$, becomes bigger as $c$ and $n$ increase. In fact,
$$
\eqalign{
\left|\Delta\E(X-c)^n \right| &= \left|\sum_{h=0}^n (-1)^h \, {n \choose h} \, c^{n-h} \, \Delta\E(X^h)\right| \cr
                                  &\leq  \sum_{h=0}^n  {n \choose h} \, c^{n-h} \, \left| \Delta\E(X^h)\right| \cr
                                  &=\parallel \Delta \, \E(X^h) \parallel_{\infty} \, \sum_{h=0}^n  \, {n \choose h} \, c^{n-h}= \cr
                                  &=\parallel \Delta \, \E(X^h) \parallel_{\infty} \, (1+c)^n \simeq eps \, (1+c)^n, \cr}
\eqno(2.6)
$$
where $eps$ corresponds to the error machine.
\par
\item{(b)} the numerical error $\Delta \, \E(X^s)$ due to the evaluation of $\E(X^s)$ involving
the first $M_{max}$ central moments $\E(X-c)^n$, is given by
$$
\eqalign{
\left|\Delta \, \E(X^s) \right| &= \sum_{n=0}^ {M_{max}} {s \choose n} c^{s-n} \Delta \,\E(X-c)^n  \cr
                                 & \leq \sum_{n=0}^{M_{max}} \left|{s \choose n}\right| c^{s-n} \left| \Delta \,\E(X-c)^n \right|  \cr
                                 & \leq {\parallel \Delta \,\E(X-c)^n \parallel}_{\infty} \,\, c^s \, \max_n{s \choose n} \, 
                                     \sum_{n=0}^{M_{max}} \left( {1 \over c} \right)^n \cr
                                 &= {\parallel \Delta \,\E(X-c)^n \parallel}_{\infty} c^s\,\,\, \max_n{s \choose n}  { {\left( {1 \over c}                                        \right)^{M_{max}+1}-1} \over {{1 \over c}-1}}, \cr}
\eqno(2.7)
$$
with $\max_n{s \choose n}= {s \choose {\left[ s/2 \right]}}$ if $[s]$ is even and
$\max_n{s \choose n}= {s \choose {\left[ s/2 \right]+1}}$
if $[s]$ is odd, where $[x]$ represents the integer part of $x$. The product of first two factors of the
right hand side of (2.7) is an increasing function of $c$, whilst the last factor gives a function which decreases with $c$.
\smallskip
\par\noindent
Hence, taking in account both (a) and (b), a reasonable choice of $c$ could be $c={1 \over 2}$. 
Further, rewriting the last  inequality in (2.7) as
$$
\left|\Delta \, \E(X)^s \right| \leq {\parallel \Delta \,\E(X-c)^n \parallel}_{\infty} \,\, c^s \, \max_n{s \choose n} \,
{ {\left( {1 \over c} \right)^{M_{max}+1}-1} \over {{1 \over c}-1}} < \varepsilon
$$
we can reconstruct the $s$-order fractional moment with a prefixed level of accuracy $\varepsilon, \, \varepsilon > 0$, 
just involving a number of central moments equal to the value $M_{max}$.
 \bigskip\noindent
3. Recovering $f(x)$ from fractional moments
\par\noindent
Let be $X$ a positive r.v. on $[0,1]$ with density $f(x)$,
Shannon-entropy $H[f]=-\int_0^1 f(x)\ln f(x)dx$ and moments
$\{\mu_j\}_{j=0}^\infty$, from which positive fractional moments
$E(X^{\alpha_j})=\sum_{n=0}^{\infty} b_n(\alpha_j) \mu_n$ may be obtained, as in (2.4)-(2.5).
\par\noindent
From [4], we know that the Shannon-entropy maximizing
density function $f_{M}(x)$, which has the same $M$ fractional moments
$E(X^{\alpha_j})$, of $f(x)$, $j=0,...,M$, is
$$
f_{M}(x)=\exp(-\sum_{j=0}^{M} \lambda_j x^{\alpha_j}).
\eqno(3.1)
$$
\noindent
Here $(\lambda_0,...,\lambda_{M})$ are Lagrangean multipliers,
which must be supplemented by the condition that the first $M$ fractional moments
of $f_M(x)$ coincide with $E(X^{\alpha_j})$, i.e,
$$
E(X^{\alpha_j})=\int_0^1 x^{\alpha_j} f_{M}(x)dx,\;j=0,...,M,
\;\;\alpha_0=1
\eqno(3.2)
$$
\noindent
The Shannon entropy $H[f_{M}]$ of $f_M(x)$ is given as
$$
H[f_{M}]=-\int_0^1 f_{M}(x)\ln f_{M}(x)dx
=\sum_{j=0}^{M} \lambda_j E(X^{\alpha_j}).
\eqno(3.3)
$$
\noindent
Given two probability densities $f(x)$ and $f_{M}(x)$, there are two
well-known measures of the distance between
$f(x)$ and $f_{M}(x)$. Namely the divergence measure
$I(f,f_{M})=\int_0^1 f(x)\ln {f(x) \over f_{M}(x)} dx$
and the variation measure
$V(f,f_{M})=\int_0^1\mid f_{M}(x)-f(x)\mid dx$.
If $f(x)$ and $f_M(x)$ have the same fractional moments
$E(X^{\alpha_j})$, $j=1,...,M$ then
$$
I(f,f_{M})=H[f_{M}]-H[f]
\eqno(3.4)
$$
\noindent
holds. In fact
$I(f,f_{M})=\int_0^1 f(x)\ln {f(x) \over f_{M}(x)} dx=
-H[f]+\sum_{j=0}^{M} \lambda_j \int_0^1 x^{\alpha_j} f_M(x) dx=
-H[f]+\sum_{j=0}^{M} \lambda_j E(X^{\alpha_j})=H[f_{M}]-H[f]$.
\par\noindent
In literature, several lower bounds for the divergence measure $I$ based
on the variation measure $V$ are available. We shall however use the following bound [7]
$$
I \geq {V^2\over 2}.\eqno(3.5)
$$
\noindent
If $g(x)$ denotes a bounded function, such that $\mid g(x)\mid\leq K$,
$K>0$, by taking into account (3.4) and (3.5), we have
$$
\mid E_f(g)-E_{f_M}(g)\mid\leq \int_0^1 \mid g(x)\mid \cdot
\mid f(x)-f_M(x)\mid dx\leq K\sqrt{2(H[f_M]-H[f])}
\eqno(3.6)
$$.
\noindent
Equation (3.6) suggests us what fractional moments have to be chosen
$$
\{\alpha_j\}_{j=1}^M:\;\; H[f_M]=\hbox{minimum}
\eqno(3.7)
$$
\noindent
The use of fractional moments in the framework of ME relies on the
following two theoretical results. The first is a theorem [8, Th. 2] which
guarantees the existence of a probability density from the knowledge
of an infinite sequence of fractional moments
\bigskip\noindent
Theorem 3.1 [8, Th. 2] If $X$ is a r.v. assuming values from a bounded interval
$[0,1]$ and $\{\alpha_j\}_{j=0}^\infty$ is an infinite sequence of positive
and distinct numbers satisfying
$\lim\limits_{j\rightarrow \infty} \alpha_j=0$ and 
$\sum_{j=0}^{\infty} \alpha_j=+\infty$, then the sequence of moments
$\{E(X^{\alpha_j})\}_{j=0}^\infty$ characterizes $X$.
\bigskip\noindent
The second concerns the convergence in entropy of $f_M(x)$,
where entropy-convergence means
$\lim\limits_{M\rightarrow \infty}H[f_M]=H[f]$.
More precisely,
\bigskip\noindent
Theorem 3.2. If $\{\alpha_j\}_{j=0}^M$ are equispaced within $[0,1)$,
with $\alpha_{M-j+1}={j\over M+1}$, $j=0,...,M$ then the ME approximant converges
in entropy to $f(x)$.
\par\noindent
Proof. See Appendix.
\bigskip\noindent
We just point out that the choice of equispaced points
$\alpha_{M-j+1}={j\over M+1}$, $j=0,...,M$ satisfies both conditions
of Theorem 3.1, i.e.
$$
\lim\limits_{M\rightarrow \infty} \alpha_M=0 \;\;\hbox{and}\;\;
\lim\limits_{M\rightarrow \infty}\sum_{j=0}^{M} \alpha_j=
\lim\limits_{M\rightarrow \infty}{1\over M+1}{M\over 2}(M+1)=+\infty.
$$
\noindent
As a consequence, if the choice of equispaced $\alpha_{M-j+1}$ guarantees
entropy-convergence, then the choice (3.7) guarantees entropy-convergence too.
\par\noindent
From a computational point of view, Lagrangean multipliers
$(\lambda_1,...,\lambda_M)$ are obtained by (1.4),
and the normalizing constant $\lambda_0$ is obtained by imposing that
the density integrates to 1.
Then the optimal $\{\alpha_j\}_{j=1}^M$ exponents are obtained as
$$
\{\alpha_j\}_{j=1}^M:\;\;\min_{\alpha_1,...,\alpha_M} \Bigl [
\min_{\lambda_1,...,\lambda_M} \Gamma(\lambda_1,...,\lambda_M) \Bigl ].
\eqno(3.8)
$$
\bigskip\noindent
4. Numerical results
\par\noindent
We compare fractional and ordinary moments by choosing some probability densities on $[0,1]$.
\par\noindent
Example 1. Let be
$$
f(x)={\pi\over 2}\sin(\pi x)
$$
\noindent
with $H[f]\simeq -0.144729886$. From $f(x)$ we have ordinary moments
satisfying the recursive relationship
$$
\mu_n={1\over 2}-{n(n-1)\over \pi^2}\mu_{n-2},\qquad n=2,3,...,
\;\; \mu_0=1,\;\; \mu_1={1\over 2}.
$$
\noindent
From $\{\mu_n\}_{n=0}^\infty$ we calculate
$E(X^{\alpha_j})=\sum_{n=0}^{\infty} b_n(\alpha_j) \mu_n$,
as in (2.4)-(2.5). From $\{E(X^{\alpha_j})\}_{j=0}^M$
we obtain the ME approximant $f_M(x)$ for increasing values of $M$,
where $\{\alpha_j\}_{j=1}^M$ satisfy (3.7).
\par\noindent
In Table 1 are reported
\par\noindent
a) $H[f_M]-H[f]=I(f,f_{M})$ and exponents $\{\alpha_j\}_{j=1}^M $ satisfying (3.7),
where $H[f_M]$ is obtained using fractional moments.
\par\noindent
b) $H[f_M]-H[f]=I(f,f_{M})$, where $H[f_M]$ is obtained using ordinary moments.
\par\noindent
Inspection of Table 1 allows us to conclude that:
\par\noindent
1) Entropy decrease is fast, so that practically 4-5 fractional
moments determine $f(x)$.
\par\noindent
2) On the converse an high number of ordinary moments are requested for a satisfactory
characterization of $f(x)$.
\par\noindent
3) Approximately 12 ordinary moments have an effect comparable to 3
fractional moments.
\par\noindent
$f(x)$ and $f_M(x)$, obtained by 4-5 fractional moments,
are practically indistinguishable.
\bigskip\noindent
\centerline{Table 1}
\par\noindent
\centerline{Optimal fractional moments and entropy difference of distributions having an}
\par\noindent
\centerline {increasing number of common a) fractional moments b) ordinary
moments}
$$
\vbox{\tabskip=0pt \offinterlineskip
\def\tablerule{\noalign{\hrule}}
\halign{\strut#&\vrule#\tabskip=1em plus 2em&
\hfil#\hfil&\vrule#&\hfil#&\vrule#&\hfil#&\vrule#
\tabskip=0pt\cr
&\multispan{7} \hfil\hfil{a)}\hfil\hfil\cr
\noalign{\smallskip}
\tablerule
&&\omit\hidewidth$M$\hidewidth&&
\omit\hidewidth $\{\alpha_j\}_{j=1}^M$\hidewidth&&
\omit\hidewidth $H[f_{M}]-H[f]$\hidewidth&\cr
\tablerule
&&$1  $&&$13.4181$&&$0.8716E-1$&\cr
&&$   $&&$       $&&$         $&\cr
&&$2  $&&$0.00289$&&$0.2938E-2$&\cr
&&$   $&&$4.69275$&&$         $&\cr
&&$   $&&$       $&&$         $&\cr
&&$3  $&&$0.04680$&&$0.3038E-3$&\cr
&&$   $&&$1.84212$&&$         $&\cr
&&$   $&&$13.2143$&&$         $&\cr
&&$   $&&$       $&&$         $&\cr
&&$4  $&&$0.00220$&&$0.3276E-4$&\cr
&&$   $&&$2.76784$&&$         $&\cr
&&$   $&&$13.7293$&&$         $&\cr
&&$   $&&$20.5183$&&$         $&\cr
&&$   $&&$       $&&$         $&\cr
&&$5  $&&$0.0024 $&&$0.1016E-4$&\cr
&&$   $&&$2.7000 $&&$         $&\cr
&&$   $&&$13.700 $&&$         $&\cr
&&$   $&&$20.500 $&&$         $&\cr
&&$   $&&$25.200 $&&$         $&\cr
\tablerule}}
\qquad\qquad
\vbox{\tabskip=0pt \offinterlineskip
\def\tablerule{\noalign{\hrule}}
\halign{\strut#&\vrule#\tabskip=1em plus 2em&
\hfil#\hfil&\vrule#&\hfil#&\vrule#
\tabskip=0pt\cr
&\multispan{5} \hfil\hfil{b)}\hfil\hfil\cr
\noalign{\smallskip}
\tablerule
&&\omit\hidewidth$ M $\hidewidth&&
\omit\hidewidth $H[f_M]-H[f]$\hidewidth&\cr
\tablerule
&&$2  $&&$0.9510E-2$&\cr
&&$   $&&$         $&\cr
&&$4  $&&$0.2098E-2$&\cr
&&$   $&&$         $&\cr
&&$6  $&&$0.7058E-3$&\cr
&&$   $&&$         $&\cr
&&$8  $&&$0.4442E-3$&\cr
&&$   $&&$         $&\cr
&&$10 $&&$0.3357E-3$&\cr
&&$   $&&$         $&\cr
&&$12 $&&$0.3288E-3$&\cr
&&$   $&&$         $&\cr
&&$   $&&$         $&\cr
&&$   $&&$         $&\cr
&&$   $&&$         $&\cr
&&$   $&&$         $&\cr
&&$   $&&$         $&\cr
&&$   $&&$         $&\cr
&&$   $&&$         $&\cr
\tablerule}}
$$
\bigskip\noindent
Example 2.
This example is borrowed from [9]. Here the authors attempt
to recover a non-negative decreasing differentiable function $f(x)$ from the
frequency moments $\omega_n$, with
$$
\omega_n=\int_0^1[f(x)]^n dx, \qquad n=1,2,...
$$
\noindent
The authors of [9] realize that other density reconstruction procedures,
alternative to ordinary moments, would be desirable. We propose fractional
moments density reconstruction procedure. Here
$$
f(x)=2\Bigl[{1\over 2}+{1\over 10}\ln({1\over Ax+B}-1)\Bigl]
\qquad B={1\over 1+e^{5}},\;\; A={1\over 1+e^{-5}}-{1\over 1+e^{5}}
$$
\noindent
with $H[f]\simeq -0.06118227$ ($f(x)$, compared to [9], contains
the normalizing constant 2). From $f(x)$ we have ordinary moments $\mu_n$
through a numerical procedure.
From $\{\mu_n\}_{n=0}^\infty$ we calculate
$E(X^{\alpha_j})=\sum_{n=0}^{\infty} b_n(\alpha_j) \mu_n$,
as in (2.4)-(2.5). Finally, from $\{E(X^{\alpha_j})\}_{j=0}^M$
we obtain the ME approximant $f_M(x)$ for increasing values of $M$,
where $\{\alpha_j\}_{j=1}^M$ satisfy (3.7).
\par\noindent
Table 2 reports:
\par\noindent
a) $H[f_M]-H[f]=I(f,f_{M})$ and exponents $\{\alpha_j\}_{j=1}^M $ satisfying
(3.7), where $H[f_M]$ is obtained using fractional moments.
\par\noindent
b) $H[f_M]-H[f]=I(f,f_{M})$, where $H[f_M]$ is obtained using ordinary moments.
\bigskip\noindent
Inspection of Table 2 allows us to conclude that:
\par\noindent
1) Entropy decrease is fast, so that practically 4 fractional
moments determine $f(x)$.
\par\noindent
2) An high number of ordinary moments is requested for a satisfactory
characterization of $f(x)$.
\par\noindent
3) Approximately 14 ordinary moments have an effect comparable to 4 fractional moments.
\par\noindent
Functions $f(x)$ and $f_M(x)$, obtained by 4 fractional moments, are practically indistinguishable.
As a consequence, we argue that the use of 4 fractional moments is as effective as that of 8 frequency
moments (as in [9]).
The former ones, indeed, provide an approximant $f_M(x)$ practically indistinguishable from $f(x)$
(see figure 1 of [9]).
\bigskip\noindent
\centerline{Table 2}
\par\noindent
\centerline{Optimal fractional moments and entropy difference of distributions
having an}
\par\noindent
\centerline {increasing number of common a) fractional moments b) ordinary
moments}
$$
\vbox{\tabskip=0pt \offinterlineskip
\def\tablerule{\noalign{\hrule}}
\halign{\strut#&\vrule#\tabskip=1em plus 2em&
\hfil#\hfil&\vrule#&\hfil#&\vrule#&\hfil#&\vrule#
\tabskip=0pt\cr
&\multispan{7} \hfil\hfil{a)}\hfil\hfil\cr
\noalign{\smallskip}
\tablerule
&&\omit\hidewidth$M$\hidewidth&&
\omit\hidewidth $\{\alpha_j\}_{j=1}^M$\hidewidth&&
\omit\hidewidth $H[f_{M}]-H[f]$\hidewidth&\cr
\tablerule
&&$1  $&&$1.56280$&&$0.6278E-2$&\cr
&&$   $&&$       $&&$         $&\cr
&&$2  $&&$0.52500$&&$0.3152E-2$&\cr
&&$   $&&$3.90000$&&$         $&\cr
&&$   $&&$       $&&$         $&\cr
&&$3  $&&$1.05000$&&$0.1169E-2$&\cr
&&$   $&&$3.00000$&&$         $&\cr
&&$   $&&$7.87500$&&$         $&\cr
&&$   $&&$       $&&$         $&\cr
&&$4  $&&$0.44062$&&$0.1025E-3$&\cr
&&$   $&&$7.65470$&&$         $&\cr
&&$   $&&$12.5262$&&$         $&\cr
&&$   $&&$63.9093$&&$         $&\cr
\tablerule}}
\qquad\qquad
\vbox{\tabskip=0pt \offinterlineskip
\def\tablerule{\noalign{\hrule}}
\halign{\strut#&\vrule#\tabskip=1em plus 2em&
\hfil#\hfil&\vrule#&\hfil#&\vrule#
\tabskip=0pt\cr
&\multispan{5} \hfil\hfil{b)}\hfil\hfil\cr
\noalign{\smallskip}
\tablerule
&&\omit\hidewidth$ M $\hidewidth&&
\omit\hidewidth $H[f_M]-H[f]$\hidewidth&\cr
\tablerule
&&$2  $&&$0.5718E-2$&\cr
&&$   $&&$         $&\cr
&&$4  $&&$0.1776E-2$&\cr
&&$   $&&$         $&\cr
&&$6  $&&$0.1320E-2$&\cr
&&$   $&&$         $&\cr
&&$8  $&&$0.6744E-3$&\cr
&&$   $&&$         $&\cr
&&$10 $&&$0.3509E-3$&\cr
&&$   $&&$         $&\cr
&&$12 $&&$0.2648E-3$&\cr
&&$   $&&$         $&\cr
&&$14 $&&$0.1914E-3$&\cr
\tablerule}}
$$
\noindent
5. Conclusions 
\par\noindent
In this paper we have faced up the Hausdorff moment problem and we have
solved it using a low number of fractional moments,
calculated explicitly in terms of given ordinary moments.
The approximating density, constrained by few fractional moments,
has been obtained by maximum-entropy method. Fractional moments
have been chosen by minimizing the entropy of the approximating density.
The strategy proposed in the present paper, for recovering a given density function, consists in
accelerating the convergence
by a proper choice of fractional moments, so obtaining an approximating
density by the use of low order moments, as (1.1) suggests.
\vfill\eject\noindent
6. References
\par\noindent
\bigskip\noindent
[1] D. Fasino, Spectral properties of Hankel matrices and numerical solutions of finite
moment problems, J. Comput. Applied Math., {\bf 65}, 145-155, (1995).
\bigskip\noindent
[2] G. Talenti, Recovering a function from a finite number of moments,
Inverse Problems, {\bf 3}, 501-517, (1987).
\bigskip\noindent
[3] S. Karlin, L.S. Shapley, {\it Geometry of moment spaces}, AMS Memoirs
{\bf 12}, Providence RI (1953).
\bigskip\noindent
[4] H.K. Kesavan, J.N. Kapur, {\it Entropy Optimization Principles with
Applications}, Academic Press, (1992).
\bigskip\noindent
[5] B. Beckermann, The condition number of real Vandermonde, Krylov
and positive definite Hankel matrices, Numerische Mathematik, {\bf 85},
553-577, (2000).
\bigskip\noindent
[6] J.M. Borwein, A.S. Lewis, Convergence of best entropy estimates,
SIAM J. Optimization, {\bf 1}, 191-205, (1991).
\bigskip\noindent
[7] S. Kullback, A lower bound for discrimination information in terms of
variation, IEEE Transaction on Information Theory, {IT-13},
126-127, 1967.
\bigskip\noindent
[8] G.D. Lin, Characterizations of Distributions via moments,
Sankhya: The Indian Journal of Statistics, {\bf 54}, Series A,
128-132, 1992.
\bigskip\noindent
[9] E. Romera, J.C. Angulo, J.S. Dehesa, The Hausdorff entropic moment
problem, J. of Math. Physics, {\bf 42}, 2309-2314, (2001).
\bigskip\noindent
[10] J.A. Shohat, J.D. Tamarkin, {\it The problem of moments},
AMS Mathematical Survey, {\bf 1}, Providence RI, (1963).
\vfill\eject\noindent
Appendix: Entropy convergence
\bigskip\noindent
A.1 Some background
\par\noindent
Let's consider a sequence of equispaced
points $\alpha_j={j \over M+1}$, $j=0,...,M$ and
$$
\mu_j=:E(X^{\alpha_j})=\int_0^1 t^{\alpha_j}f_M(t)dt,\qquad j=0,...,M
\eqno(A.1)
$$
\noindent
with $f_M(t)=\exp(-\sum_{j=0}^M \lambda_j t^{\alpha_j})$.
With a simple change of variable $x=t^{1 \over M+1}$, from (A.1) we have
$$
\mu_j=E(X^{\alpha_j})=\int_0^1 x^j \exp\Bigl [
-(\lambda_0-\ln (M+1))-\sum_{j=1}^M \lambda_j x^j+
M\ln x\Bigl ]dx,\;\; j=0,...,M
\eqno(A.2)
$$
\noindent
which is a reduced Hausdorff moment problem for each fixed $M$ value and
a determinate Hausdorff moment problem when $M\rightarrow\infty$.
Referring to (A.2) the following symmetric definite positive Hankel matrices
are considered
$$
\Delta_0=\mu_0,\;\;
\Delta_{2}=
 \bmatrix
          \mu_0  &\mu_{1}  \\
          \mu_{1}&\mu_{2}
 \endbmatrix,...,
\Delta_{2M}=
 \bmatrix
          \mu_0&\cdots&\mu_{M}  \\
         \vdots&\cdots&\vdots  \\
          \mu_{M}&\cdots&\mu_{2M}
 \endbmatrix
\eqno(A.3)
$$
\noindent
whose $(i,j)$-th entry $i,j=0,1,...$ holds
$$
\mu_{i+j}=\int_0^1 x^{i+j} f_M(x)dx,
$$
\noindent
where $f_M(x)=\exp\Bigl [
-(\lambda_0-\ln (M+1))-\sum_{j=1}^M \lambda_j x^j+
M\ln x\Bigl ]$.
The Hausdorff moment problem is determinate and the underlying distribution
has a continuous distribution function $F(x)$, with density $f(x)$.
Then the massimal mass $\rho (x)$ which can be concentrated at any real
point $x$ is equal to zero ([10], Corollary (2.8)).
In particular, at $x=0$ we have
$$
0=\rho (0)=
\lim\limits_{i\rightarrow \infty} \rho_i^{(0)}=:
         {\mid \Delta_{2i}\mid \over
 \vmatrix
          \mu_2&\cdots&\mu_{i+1}  \\
         \vdots&\cdots&\vdots  \\
          \mu_{i+1}&\cdots&\mu_{2i}
 \endvmatrix } =
\lim\limits_{i\rightarrow \infty} (\mu_0-\mu_0^{-(i)})
\eqno(A.4)
$$
\noindent
where $\rho_i^{(0)}$ indicates the largest mass which can be concentrated
at a given point $x=0$ by any solution of a reduced moment problem
of order $\geq i$ and $\mu_0^{-(i)}$ indicates the minimum value
of $\mu_0$ once assigned the first $2i$ moments.
\par\noindent
Let's fix
$\{\mu_0,...,\mu_{i-1},\mu_{i+1},...,\mu_{M}\}$ while only
$\mu_i$, $i=0,...,M$ varies continuously. From (A.2) we have
$$
\Delta_{2M}\cdot
             \bmatrix d\lambda_0/d\mu_i     \\
                                 \vdots     \\
                      d\lambda_{M}/d\mu_i
             \endbmatrix
=-e_{i+1}
\eqno(A.5)
$$
\noindent
where $e_{i+1}$ is the canonical unit vector $\in I\!\!R^{M+1}$, from which
$$
0<\Bigl [{d\lambda_0\over d\mu_i},...,{d\lambda_M\over
d\mu_i}\Bigl ]\cdot\Delta_{2M}\cdot
             \bmatrix d\lambda_0/d\mu_i     \\
                                 \vdots     \\
                      d\lambda_{M}/d\mu_i
             \endbmatrix
=-\Bigl [{d\lambda_0 \over d\mu_i},...,{d\lambda_M \over d\mu_i}\Bigl ]
e_{i+1}= -{d\lambda_i \over d\mu_i} \qquad \forall i
\eqno(A.6)
$$
\noindent
A.2 Entropy convergence
\par\noindent
The following theorem holds.
\par\noindent
Theorem A.1
If $\alpha_j={j \over M+1}$, $j=0,...,M$ and
$f_M(x)=\exp(-\sum_{j=0}^M \lambda_j x^{\alpha_j})$ then
$$
\lim\limits_{M\rightarrow \infty}
H[f_M]=:-\int_0^1 f_M(x)\ln f_M(x) dx=
H[f]=:-\int_0^1 f(x)\ln f(x) dx.
\eqno(A.7)
$$
\noindent
Proof. From (A.1) and (A.7) we have
$$
H[f_M]=\sum_{j=0}^M \lambda_j \mu_j
\eqno(A.8)
$$
\noindent
Let's consider (A.8). When only $\mu_0$ varies
continuously, taking into account (A.3)-(A.6) and (A.8) we have
$$
{d\over d\mu_0}H[f_M]=\sum_{j=0}^M\mu_j {d\lambda_j\over d\mu_0}+
\lambda_0=\lambda_0-1
$$
$$
{d^2\over d\mu_0^2}H[f_M]={d\lambda_0\over d\mu_0}=-
{ \vmatrix
          \mu_2&\cdots&\mu_{M+1}  \\
         \vdots&\cdots&\vdots  \\
          \mu_{M+1}&\cdots&\mu_{2M}
 \endvmatrix \over \mid\Delta_{2M}\mid}=-{1\over \mu_0-\mu_0^{-(M)}}<0.
$$
\noindent
Thus $H[f_M]$ is a concave differentiable function of $\mu_0$.
When $\mu_0\rightarrow \mu_0^{-(M)}$ then
$H[f_M]\rightarrow -\infty$, whilst at $\mu_0$ it holds
$H[f_M] > H[f]$, being $f_M(x)$ the maximum
entropy density once assigned $(\mu_0,...,\mu_M)$.
Besides, when $M\rightarrow\infty$ then $\mu_0^{-(M)}\rightarrow\mu_0$.
So the theorem is proved.
\vfill\eject\noindent
\centerline{\bf HAUSDORFF MOMENT PROBLEM VIA FRACTIONAL MOMENTS}
\bigskip\noindent
\centerline{Pierluigi Novi Inverardi$^{(1)}$, Alberto Petri$^{(2)}$, Giorgio Pontuale$^{(2)}$,
Aldo Tagliani$^{(1)(*)}$}
\bigskip\noindent
$^{(1)}$ Faculty of Economics, Trento University, 38100 Trento, Italy.
\par\noindent
$^{(2)}$ CNR, Istituto di Acustica "O.M. Corbino", 00133 Roma, Italy.
\par\noindent
$^{(*)}$ Corresponding author:
\par\noindent
Phone: +39-0461-882116, Fax:+39-0461-882124, E-mail: ataglian\@cs.unitn.it
\bigskip\noindent
Abstract
\par\noindent
We outline an efficient method for the reconstruction of a probability
density function from the knowledge of its infinite sequence of
ordinary moments. The approximate density is obtained resorting to maximum
entropy technique, under the constraint of some fractional moments.
The latter ones are obtained explicitly in terms of the infinite sequence
of given ordinary moments. It is proved that the approximate density
converges in entropy to the underlying density, so that it demonstrates to be
useful for calculating expected values.
\vfill\bye